\documentclass[
11pt,tightenlines,nofootinbib,preprintnumbers,showpacs,prd]{revtex4}
\usepackage{graphicx}

\begin{document}
\preprint{CPT-2003/P.4526}
\title{Statistical Approach for Unpolarized Fragmentation Functions\\ for 
the Octet Baryons}
\author{ Claude Bourrely
\footnote{Electronic address: Claude.Bourrely@cpt.univ-mrs.fr}
and Jacques Soffer
\footnote{Electronic address: Jacques.Soffer@cpt.univ-mrs.fr}}
\affiliation{Centre de Physique Th\'eorique\footnote{Unit\'e propre 
de  Recherche 7061}, 
CNRS Luminy case 907,\\
F-13288 Marseille Cedex 09, France}

\begin{abstract} 
A statistical model for the parton distributions in the nucleon has proven its
efficiency in the analysis of deep inelastic scattering data, so we propose 
to extend this approach to the description of unpolarized fragmentation 
functions for the octet baryons. 
The characteristics of the model are determined by using some data
on the inclusive production of proton and $\Lambda$ in unpolarized deep 
inelastic scattering and a next-to-leading analysis of the available 
experimental data on the production of unpolarized octet baryons 
in $e^+e^-$ annihilation. 
Our results show that both parton distributions and fragmentation 
functions are compatible with the statistical approach, in terms of a few free
parameters, whose interpretation will be discussed.
\end{abstract}

\maketitle

\section{Introduction}

Following the first evidence of partonic substructure of the nucleon,  
by means of deep inelastic scattering (DIS), a large amount 
of experimental data have been collected in order to understand the parton 
structure of the nucleon. It is currently known that 
the $Q^2$ evolution behavior of the parton distribution functions (PDF)
is well described by perturbative QCD. However, the
parton distributions at an initial scale reflect the nonperturbative 
quark and gluon dynamics of QCD bound states, which cannot be 
determined from first principles. For this reason, many  
parametrizations have been proposed 
~\cite{GRVupol,MRSTupol,CTEQ5,GRSVpol,GSpol}, but 
most of these expressions involve a large number of free parameters, with  
no clear physical meaning. However, some
efforts have been recently made~\cite{BSB01} to parametrize 
unpolarized and polarized nucleon PDF  based on a statistical 
approach~\cite{BSb,Bha}. 
The nucleon PDF, which involve in this framework a small number of free 
parameters, can well describe all available experimental unpolarized 
and polarized DIS data, so this is perhaps an indication
that the PDF retain some important statistical features of the nucleon. 
The statistical approach of the nucleon PDF allows to make predictions, 
which were tested recently in a satisfactory way, by various experimental 
data in DIS \cite{BSB02}. In order to extend further this framework to new 
areas, it is natural to envisage, for example, its application 
to the description of the PDF of the other octet baryons. 
Unfortunately, these PDF are not directly accessible because, due to their 
short lifetimes, the hyperons cannot be used as a target in a DIS experiment.

However, it is well known that the quark distributions in a hadron 
$h$, $q_h(x,Q^2)$ are related to the corresponding quark to hadron 
fragmentation function (FF) $D_q^h(x,Q^2)$, by means of the  
so called Gribov-Lipatov relations~\cite{GLR}. This connection between two 
basic quantities of the hadron structure has been used in several 
recent works \cite{MSSY,MSSY02}, as an attempt to improve our present poor 
knowledge on the hadron FF.
So this is our motivation to extend the statistical approach to a global 
description of the octet baryon FF and to check its validity 
against the available experimental data. 

The paper is organized as follows. In Sect. 2, we review the main points of 
the framework and we give the basic procedure for the construction of the
octet baryons FF in the statistical approach, in terms of Fermi-Dirac 
distributions. In Sect. 3, we determine all free parameters of the model 
by using some data on the inclusive production of proton and $\Lambda$ 
in unpolarized deep inelastic scattering and a next-to-leading (NLO) 
fit to the available experimental data on the production of unpolarized octet 
baryons in $e^+e^-$ annihilation. The results of the fit for the
cross sections and the obtained FF are also presented and we compare our
approach with some previous works on baryon FF 
\cite{Florian, Thomas, IMR, JJY}. Finally in Sec. 4, we give our conclusions.

\section{The statistical approach for unpolarized FF}   

In the statistical approach, a hadron can be viewed as a gas of massless 
partons (quarks, antiquarks, gluons) in equilibrium at a 
given temperature, in a finite size volume. In hadron production,
when a parton fragments into a hadron, it picks up other 
partons from the QCD vacuum in order to form a specific hadron.
The formation probability of the hadron is characterized by its statistical 
properties. Therefore we believe that statistical features which were proposed
to build up the nucleon PDF in Ref.~\cite{BSB01}, can be used also 
to construct the FF for the octet baryons.
We assume that the parton ($p$) to hadron ($h$) FF $D_{p}^h(x)$, 
at an input energy scale $Q_0^2$, is proportional to

\begin{equation}
[\exp[(x - X_{0})/{\bar x}] \pm 1]^{-1}~,
\label{1}
\end{equation}
where the {\it plus} sign for quarks and antiquarks, corresponds to a 
Fermi-Dirac distribution and the {\it minus} sign for gluons, corresponds 
to a Bose-Einstein distribution. Here $X_{0}$ is a constant which plays 
the role of the {\it thermodynamical potential} of the quark 
hadronization into a hadron and $\bar x$ is the 
{\it universal temperature}, which is assumed to be equal 
for all octet baryons. It is reasonable to take its value to be the same as 
for the nucleon PDF, i.e. $\bar x=0.099$, according to Ref.~\cite{BSB01}.  
The statistical approach for the PDF allows to construct quark distributions
of a given helicity and it is also possible to relate simply quark to
antiquark distributions, resulting from chiral properties of QCD, as explained 
in Ref.~\cite{BSB01}. All these physical quantities were determined 
precisely due to the existence of a hudge amount of data in unpolarized and
polarized DIS. In the case of the octect baryon FF, the
situation is different and the sarcacity of the polarized data does not allow 
such a clear separation, so we will restrict ourselves to the determination
of the unpolarized quark FF, although the extention to the polarized case can
be easily done. For the quarks $q=u,s,d$ the FF are then expressed as

\begin{equation}
D_{q}^{B}(x, Q^2_0)= {A_{q}^{B} X_{q}^{B} {x^{b}}
\over \exp[ (x - X_{q}^{B})/ 
\bar{x}] + 1 },
\label{2}
\end{equation}
where $X_q^B$ is the potential corresponding to the fragmentation 
$q \rightarrow B$ and $Q_0^2$ is an initial scale, given below in Table 1.
We will ignore the antiquark FF $D_{\bar q}^B$, which are considered to
be strongly suppressed. 
The heavy quark FF $D^B_Q(x, Q_0^2)$ for $Q = c, b, t$, which are
expected to be large only in the small $x$ region ($x \leq 0.1$ or so), 
are parametrized by a diffractive term with a vanishing potential   
\begin{equation}
D_{Q}^{B}(x, Q_0^2)= {\tilde{A}_{Q}^B x^{\tilde{b}} \over 
\exp(x/ \bar {x}) +1}.
\label{3}
\end{equation}
The initial scale $Q_0^2$, which is flavor dependent in this case, is given 
below in Table 1
\footnote{ Due to the fact that the input scale of the $t$ quark is above the 
highest energy data investigated in this work, it does not contribute 
to our analysis.}.
This FF for $Q \rightarrow B$ depends on $ \tilde b$ and a
normalization constant $\tilde A _B^Q$ for each baryon $B$. For the other 
quarks, we make some reasonable assumptions in order to 
reduce the number of parameters in addition to $b$, the universal power 
of $x$ in Eq. (\ref{2}). First we have the obvious constraints, 
namely, $D_u^B =D_d^B$ for $B=p,\Lambda$. Moreover we assume that we need 
only {\it four} potentials, two for the proton $X_u^p = X_d^p$ and
$X_s^p$ and two for the hyperons $X_u^Y = X_d^Y$ and $X_s^Y$ where 
$Y = \Lambda, \Sigma^{\pm}, \Xi^-$. Finally for the gluon to baryon FF
$D_g^B(x,Q^2)$, which is hard to determine precisely, we take a Bose-Einstein 
expression with a vanishing potential
\begin{equation}
D_{g}^{B}(x, Q_0^2)= {A_{g}^B x^{\tilde{b}+1} \over 
\exp(x/ \bar {x}) - 1}.
\label{4}
\end{equation}
We assume it has the same small $x$ behavior as the heavy quarks and 
it is the same for all baryons. 
The normalization constants $A_q^B$, $A_g^B$ and $\tilde {A}_Q^B$
will have also to be determined by fitting the data, a procedure we present 
now.

\section{Determination of the parameters from the data analysis}

The experimental data which was used is twofold. First, hadron production 
in DIS gives access to a direct determination of $D_u^p(x,Q^2)$ 
\cite{EMC89} and $D_u^{\Lambda}(x,Q^2)$ \cite{HERMES00}, in a limited range 
of $x$ and $Q^2$. 
We have also determined the free parameters of the model by making an analysis 
of the differential cross section for the semi-inclusive hadron production 
process $e^+e^- \to h + X$, for $p, \Lambda, \Sigma,\Xi$. Here we note
that in these experiments, they usually do not distinguish between $B$ and 
$\bar B$, so we will include both contributions in our calculations 
by making the natural assumption $D_q^B(x,Q^2)=D_{\bar q}^{\bar B}(x,Q^2)$.
The differential cross section can be expressed as
\cite{Altarel,Furman,Kretzer,Binnew}

\begin{equation}
\frac{1}{\sigma_{tot}}\frac{d \sigma}{d x_E} = \sum\limits_q
\int\limits_{x_E}^1 \frac{d\eta}{\eta}
D_q^{h}(\frac{x_E}{\eta}, \mu^2_F) \frac{1}
{\sigma_{tot}}\left[C^q_T(\eta,Q^2,\mu^2_{F,R})+
C^q_L(\eta,Q^2,\mu^2_{F,R})\right]\,,
\label{sigmatot}
\end{equation}
where $x_E=2 E_{h}/\sqrt{s}$, which is the energy $E_h$ of the produced hadron
scaled to the beam energy $Q/2 \equiv \sqrt{s}/2$.
\footnote{When experimental data are presented in the momentum scaling
variable $z_p = 2p_h/Q$, we made a conversion to $x_E$ in the following
figures.}
The subscripts $T$ and $L$ denote the contributions due to transverse and 
longitudinal polarizations, respectively. The summation $q$ includes
quarks, antiquarks and gluon contributions. In the following we will set
the renormalization and the factorization scales equal to
$\mu^2_R = \mu^2_F = Q^2$. The definitions of the functions $C^q_T$ and
$C^q_L$ follow those of Ref. \cite{Kretzer}.
In Eq. (\ref{sigmatot}), $\sigma_{tot}$ is the total cross section 
for the process.
\begin{equation}
\label{sigtot}
\sigma_{tot}=N_c\sum_{q}\frac{4 \pi \alpha^2}{3 s} 
\hat{e}_{q}^2(s) \left(1 +\frac{\alpha_s(s)}{\pi}\right)\,,
\end{equation}
where $N_c$ is the color number and we shall take $N_c=3$,
$\alpha$ is the QED fine structure constant and $\alpha_s(s)$ is 
the strong coupling constant. The sum in the above equation
should be over all active quarks and antiquarks.
The electroweak charges in Eq. (\ref{sigtot}) can be expressed as
\begin{equation}
\hat{e}_q^2(s)=e_q^2+2 v_e v_q e_q\rho_1(s) +(a_e^2+v_e^2)
(a_q^2+v_q^2)\rho_2(s)\,,
\label{hatC}
\end{equation}
with
\begin{equation}
\rho_1(s)=\frac{1}{4 \sin^2 \theta_W \cos^2 \theta_W}
\frac{s(M_Z^2-s)}{(M_Z^2-s)^2+M_Z^2\Gamma_Z^2}\,,
\label{rho1}
\end{equation}

\begin{equation}
\rho_2(s)=\frac{1}{16 \sin^4 \theta_W \cos^4 \theta_W}
\frac{s^2}{(M_Z^2-s)^2+M_Z^2\Gamma_Z^2}\,,
\label{rho2}
\end{equation}

\begin{equation}
a_e=-1,~~a_q = T_{3q}\,,
\label{cplax}
\end{equation}
\begin{equation}
v_e=-\frac{1}{2}+2 \sin^2 \theta_W,~~~~v_q=T_{3q}-2e_q \sin^2 \theta_W\,,
\label{cplvec}
\end{equation}
where $e_q$ is the electric charge of the quarks and ($v_i$), 
($a_i$) the electroweak vector and axial couplings of the electron and
the quarks to the $Z$, respectively. $T_{3q}$ is the third component of
the strong isospin, $T_{3q}=1/2,-1/2$ for up-type quark and down-type quark, 
respectively. $\theta_W$ is the weak-mixing angle, and $M_Z$, $\Gamma_Z$ 
are the mass and width of the $Z$. 

The FF have been evolved at NLO following a method defined in Ref. 
\cite{BSB01}, where for the DGLAP equations  we used the 
{\it timelike} splitting functions calculated in Refs. \cite{DGLAP}. 
For baryon production in $e^+e^-$ collisions, we made a NLO fit with 
formula (\ref{sigmatot}) of the experimental data from 
Refs.~\cite{TASSO83}-\cite{DELPHI95b}, where we restricted $x_E \ge 0.1$.
In addition, the input scales choice \cite{Kretzer} and  
$\Lambda ({\overline{MS}})$ are given in Table \ref{table1}.

\begin{table}[htb]
\caption{Input scale $Q_0$ and  $\Lambda ({\overline{MS}})$ in GeV unit.}
\label{table1}
\begin{ruledtabular}
\begin{tabular}{ c c c c c }
quark & u,d,s & c & b & t   \\
\hline 
$Q_0$ & 0.632 & 1.4 & 4.5 & 175\\
$\Lambda ({\overline{MS}})$ & 0.299 
& 0.246 & 0.168 & 0.068
\end{tabular}
\end{ruledtabular}
\end{table}

Now, let us report the values of the free parameters we have obtained
from the NLO fit:
\begin{equation}
\begin{array}{cccc}
~~~X_u^{p}= 0.648, &  X_{s}^{p}= 0.247, &  X_u^{\Lambda}= 0.296, 
&  X_{s}^{\Lambda}= 0.476 \\ 
~~~b= 0.200, & \tilde{b}= -0.472, &  A_g^B= 0.051.
\end{array}
\label{potent}
\end{equation}
These parameters have similar values to those obtained for the
nucleon PDF \cite{BSB01}, in particular, for the thermodynamical
potentials, so the intrinsic properties of the quarks when observed
in DIS or in fragmentation processes seem to be preserved.
Notice that in the nucleon PDF the $u$ quark which is dominant has the larger
potential and here we have analogously,  $X_u^p > X_{s}^{p}$ and 
$X_{s}^{\Lambda}>X_u^{\Lambda}$, a situation which is natural to expect.
The other  parameters $A_{q_1}^{B}$, $A_{q_2}^{B}$, for the quark to 
baryon FF, and $\tilde{A}_{Q}^B$ for the heavy quarks are given 
in Table~\ref{table2}, together with the quark content
$q_1$, $q_2$ and $Q$ for each baryon $B$. Clearly these normalization 
constants are decreasing when going from the proton to the heavier hyperons, 
following the magnitude of the corresponding measured cross sections.

\begin{table}[ht]
\caption{Values of the normalization constants of the FF
for the octet baryons}
\label{table2}
\begin{ruledtabular}
\begin{tabular}{ c c c c c c }
\hline
\hline
Baryon & $q_1$& $q_2$ & $A_{q_1}^{B}$ & $A_{ q_2}^{B}$&  $\tilde{A}_{Q}^B$
\\ \hline
 ~~~~$p(uud)$~~~~ & $~~ u=d ~~$  &$~~ s ~~$
& 0.264 & 1.168 & 2.943  
 \\
 ~~~~$\Lambda(uds)$~~~~ & $~~ u=d ~~$ & $~~ s ~~$ 
& 0.428 & 1.094 & 0.720 
 \\ 
 ~~~~$\Sigma^+(uus)$~~~~ & $~~ u ~~$ & $~~ s ~~$ 
& 0.033 & 0.462 & 0.180  
 \\ 
 ~~~~$\Sigma^-(dds)$~~~~ & $~~ d ~~$ & $~~ s ~~$ 
& 0.030 & 0.319 & 0.180  
 \\ 
 ~~~~$\Xi^-(dss)$~~~~ & $~~ d ~~$ & $~~ s ~~$ 
& 0.023 & 0.082 & 0.072   
 \end{tabular}
\end{ruledtabular}
\end{table}

Finally, let us comment on our results. With  the data set on proton, 
$\Lambda$, $\Sigma^+$, $\Sigma^-$ and $\Xi^-$ production, we get a 
$\chi^2 = 227.5$ for 206 experimental points ($x_{E} \geq 0.1$), 
{\it i.e.} $\chi^2/\mbox{point}=1.1$ for our NLO fit. 
This agreement of our results with experimental data 
is satisfactory and it confirms that the statistical approach 
is also successful in the description of the octet baryons FF.  

\begin{figure}[htb]
\centerline{\includegraphics[bb=50 100 584 754,width=12cm]{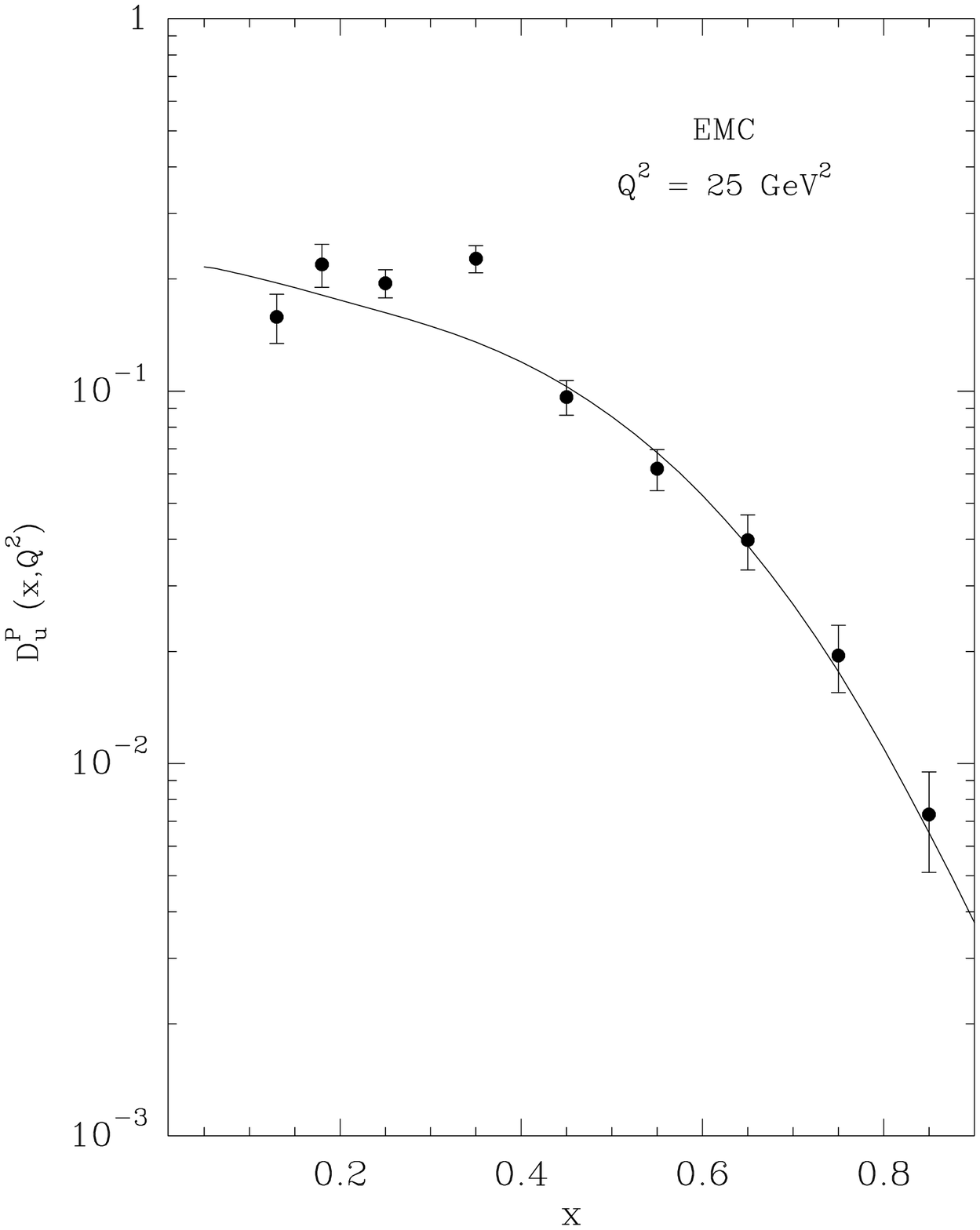}}
\caption[*]{\baselineskip 13pt
The $u$ quark to proton FF $D_u^p(x,Q^2)$ as a function of $x$ 
at $Q^2 = 25\rm{GeV^2}$. 
The experimental data are from Ref. \cite{EMC89}.}
\label{du-proton}
\end{figure}

\begin{figure}[htb]
\centerline{\includegraphics[bb=50 100 584 754,width=12cm]{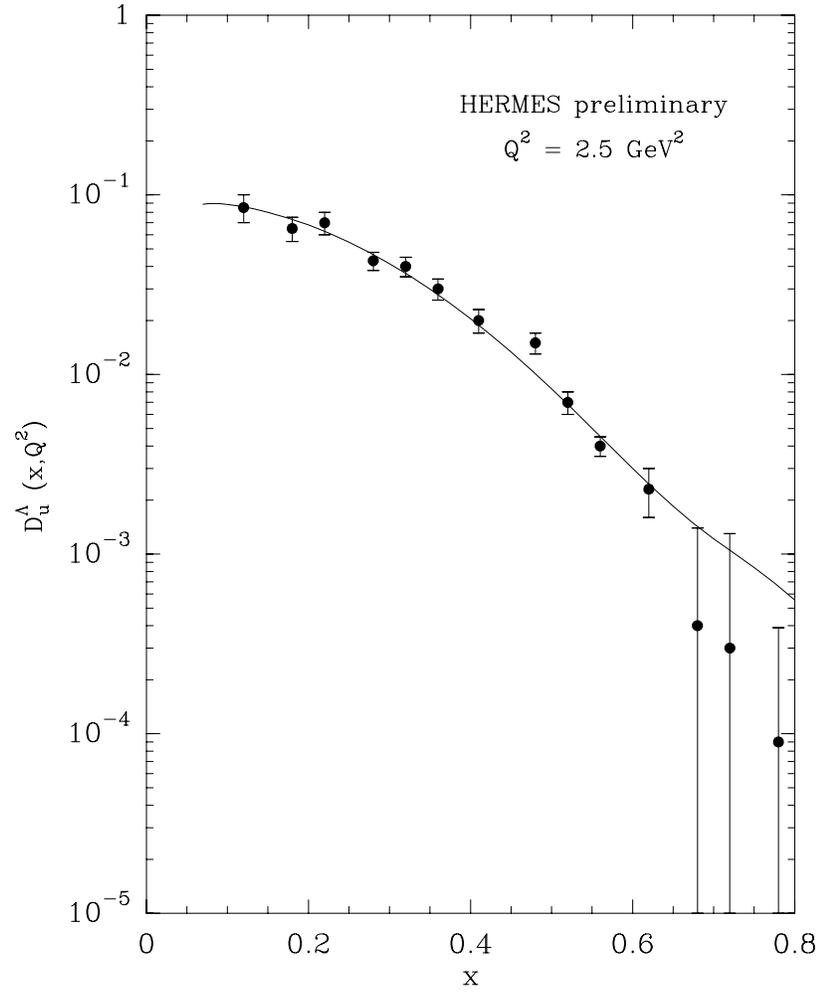}}
\caption[*]{\baselineskip 13pt 
The FF for $u$ quark to $\Lambda$, $D_u^{\Lambda}(x,Q^2)$, 
as a function of $x$ at $Q^2 = 2.5\rm{GeV^2}$. 
The experimental data are from Ref. \cite{HERMES00}.}
\label{du-lambda}
\end{figure}

\begin{figure}[ht]
\centerline{\includegraphics[bb=50 100 584 754,width=12cm]{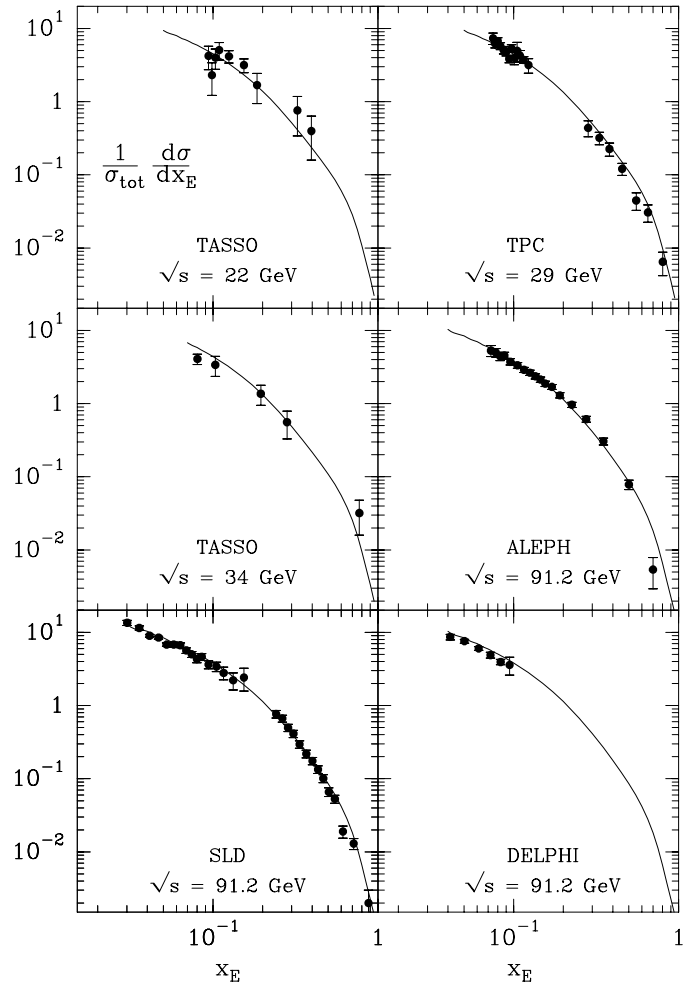}}
\caption[*]{\baselineskip 13pt 
Cross sections for proton production in $e^+e^-$ annihilation at several 
energies as function of $x_E$. The experimental data 
are from Refs.~\cite{TASSO83,TPC88,TASSO89,ALEPH95,DELPHI95a,SLD99}.}
\label{sigproton}
\end{figure}

\begin{figure}[htb]
\centerline{\includegraphics[bb=50 100 584 754,width=13cm]{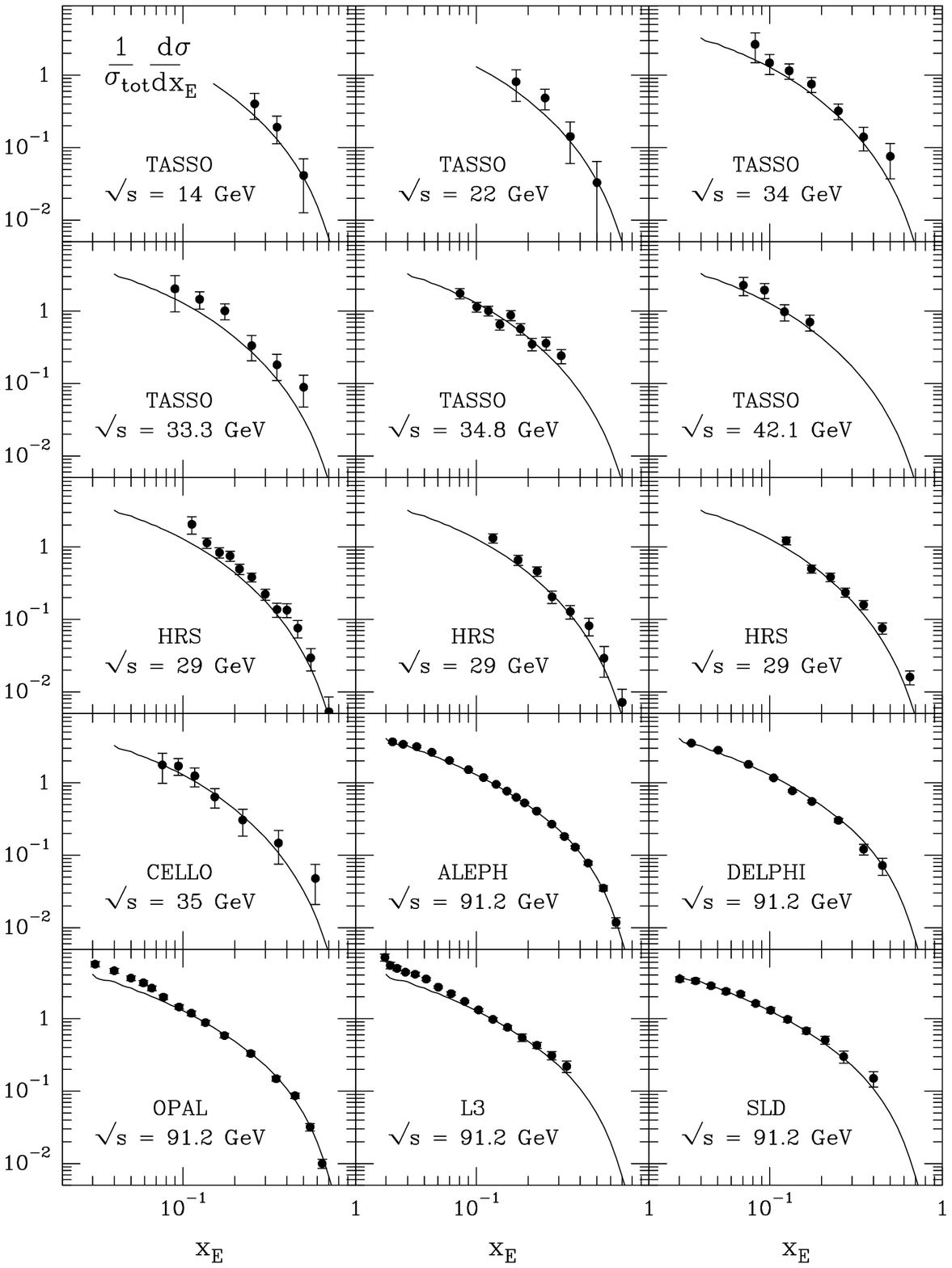}}
\caption[*]{\baselineskip 13pt 
Cross sections for $\Lambda$ production in $e^+e^-$ annihilation at several
energies, as function of $x_E$. The experimental data are from 
Refs.~\cite{SLD99,TASSO81,TASSO85,TASSO90,HRS86,HRS87,
HRS92,CELLO90,ALEPH98,DELPHI93,OPAL96,L3}.}
\label{siglam}
\end{figure}

\begin{figure}[htb]
\centerline{\includegraphics[bb=50 100 584 754,width=12cm]{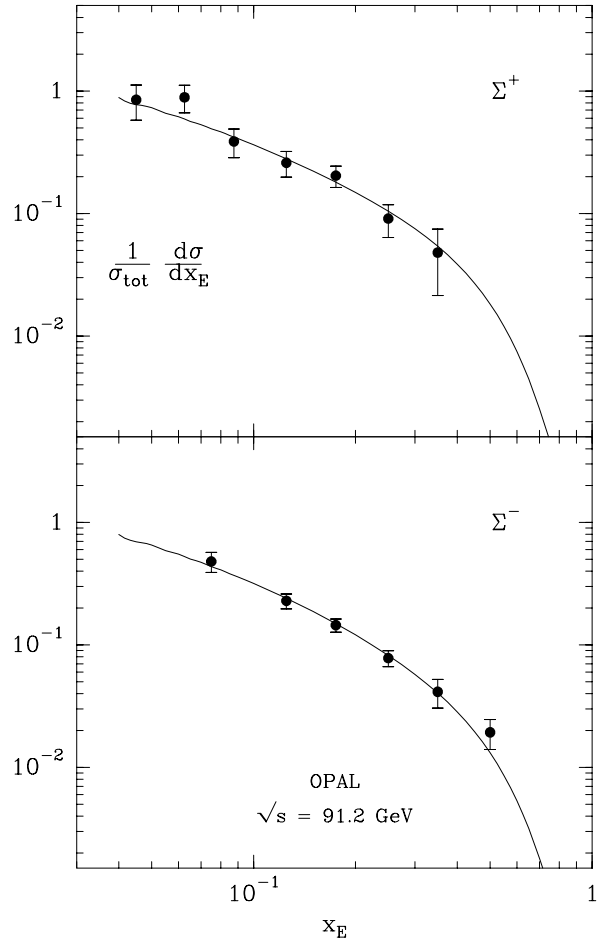}}
\caption[*]{\baselineskip 13pt
Cross sections for  $\Sigma^\pm$ production in $e^+e^-$ annihilation at the 
Z-pole as function of $x_E$.
The experimental data are from Ref. \cite{OPAL97}.}
\label{sigsigma}
\end{figure}

\begin{figure}[htb]
\centerline{\includegraphics[bb=50 100 584 754,width=12cm]{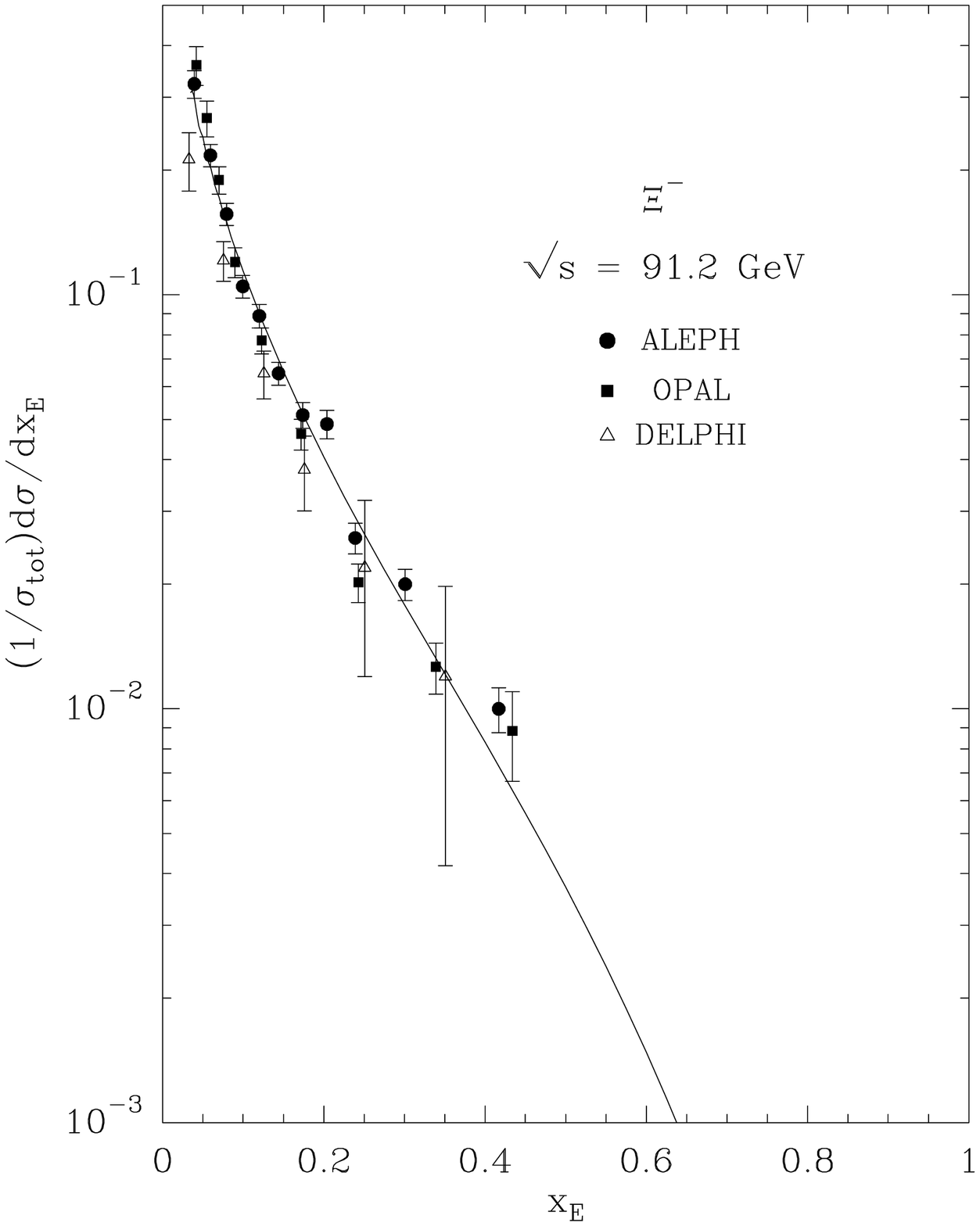}}
\caption[*]{\baselineskip 13pt 
Cross sections for $\Xi^-$ production in $e^+e^-$ annihilation at the 
Z-pole as function of $x_E$.
The experimental data are from Refs. \cite{ALEPH98,OPAL96,DELPHI95b}.}
\label{sigxi}
\end{figure}

In Figs. \ref{du-proton} and \ref{du-lambda} we display the available DIS
data, which yield directly $D_u^p(x,Q^2)$ and $D_u^{\Lambda}(x,Q^2)$, and the 
result of our fit. 
In Figs. \ref{sigproton}-\ref{sigxi} we show a comparison of the calculated
cross sections for various baryons with experimental data.
In Fig. \ref{sigproton}, we give our results for the proton cross section 
in electron-positron annihilation at $\sqrt{s}=22, 29, 34~\rm{GeV}$
and at the Z-pole {\it i.e.} $\sqrt{s}= 91.2~\rm{GeV}$.
The agreement with data is very good even in the low $x_{E}$ region 
($x_{E} \leq 0.1$) as shown on the same figure, although these data points
were not included in our fit.

\begin{figure}[htb]
\centerline{\includegraphics[bb=50 100 584 754,width=12cm]{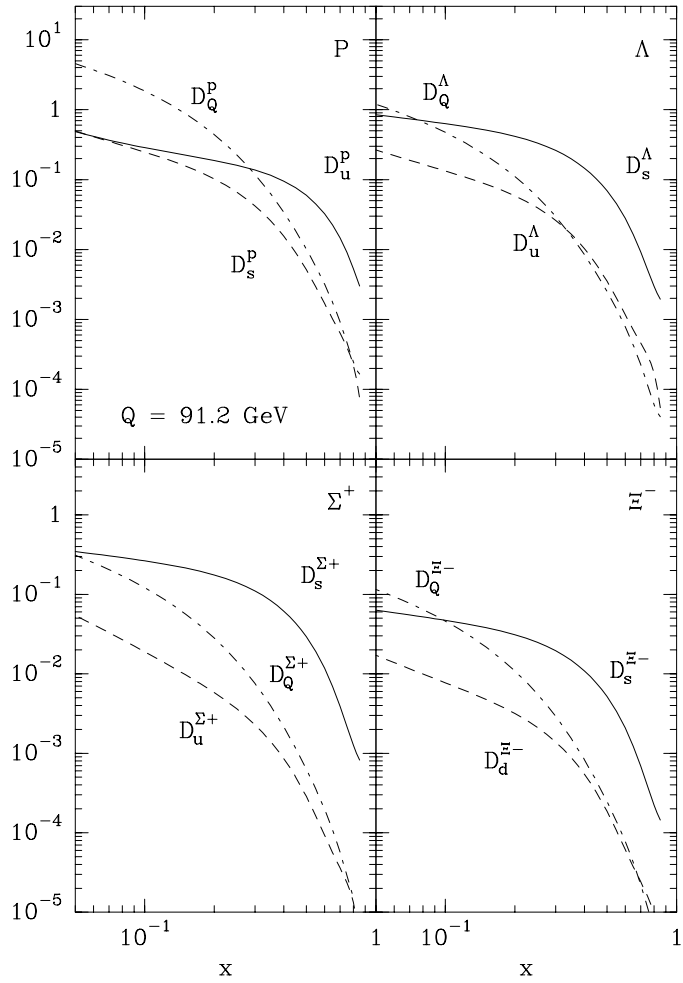}}
\caption[*]{\baselineskip 13pt
The quark to octet baryons FF $D_q^B(x,Q^2)$ and  $D_Q^B(x,Q^2)$ 
($B = p,\Lambda, \Sigma^{\pm}, \Xi^-$ , $q = u, d, s$  and $Q = c, b, t$), as 
a function of $x$ at  $Q = 91.2\rm{GeV}$. Note that we used different vertical 
scales in the upper and lower parts of the figure.}
\label{fragoctet} 
\end{figure}

For the $\Lambda$ production, our results are displayed in Fig. \ref{siglam}
for the energies $\sqrt{s}=14, 22, 29, 33.3, 34.8, 42.1, 91.2~\rm{GeV}$.
It is clear from Figs.~\ref{sigproton} and \ref{siglam}, which display a
sizeable energy domain, that the scaling violations in this range are very 
small.
In order to improve the small $x_{E}$ behavior, we need to modify
the evolution of the FF in this region, and also include finite mass 
corrections and modifications of the splitting functions 
(see a discussion in Ref. \cite{Florian}), but such corrections are outside 
the scope of the present paper. 
In Fig.~\ref{sigsigma} we give our results for $\Sigma^\pm$ production at 
$\sqrt{s}=91.2~\rm{GeV}$ and in Fig.~\ref{sigxi} our results for $\Xi^-$ 
production at the Z-pole. 
For these strangeness production processes, our results show a satisfactory 
agreement. The corresponding quark to baryon FF are presented in 
Fig.~\ref{fragoctet}. For all hyperons the strange quark FF dominates largely 
over the $u,d$ quarks, which seems natural. For $\Lambda$, we are in agreement 
with a model for SU(3) symmetry breaking in Ref. \cite{IMR}, which leads to
$D_u^{\Lambda} \sim 0.07 D_s^{\Lambda}$. In Ref. \cite{Thomas} one also finds 
$D_u^{\Lambda}$ smaller than $D_s^{\Lambda}$ and $D_{\bar q}^{\Lambda}$ is
strongly suppressed. This is in contrast with the situation of 
Ref. \cite{Florian}, where $u,d$ and $s$ are assumed to contribute equally. 
However the heavy quarks have a pattern similar to Ref. \cite{Florian}, 
with a sizeable contribution only for $x \leq 0.1$ and a fast dropping off for 
large $x$. This is also at variance with Ref. \cite{JJY}, where  
$D_u^{\Lambda} / D_s^{\Lambda}$ decreases from 1 to 0.2 when $x$ goes from 0 
to 1. 
For the proton it is surprizing to see that the $u$-quark FF dominates only 
at large $x$, whereas the strange and heavy quarks contribute substantially 
for $x \leq 0.3$ or so. Finally we notice that for the $\Lambda$, the heavy 
quarks have a pattern similar to Ref. \cite{Florian}, with a sizeable 
contribution only for $x \leq 0.1$ and a fast dropping off for large $x$.

\section{Conclusions}

With the motivation to check whether the octet baryons FF have similar 
statistical features as the nucleon PDF previously studied, we extended the 
statistical approach to analyze some data on the inclusive production of the 
octet baryons.  
We found that these FF can be well described with a small number of 
free parameters, whose interpretion was discussed. 
We obtained a satisfactory description of the unpolarized experimental data, 
suggesting that the statistical approach, with Fermi-Dirac type FF, 
works equally well, compared to other parametrization forms.
The semi-inclusive DIS data for proton and $\Lambda$ production, which give 
strong constraints, have allowed a flavor separation between $u,d$ and $s$
quarks FF, an interesting result which remains to be more seriously checked 
in the future.

In our present analysis, we did not introduce polarized FF, although our 
formalism can be easily extended to this case.
Actually, it is impossible to extract some reliable information 
on the polarized FF for the moment, due to the scarcity of the data, but
we hope this will be possible in the future.

\begin{acknowledgments}
We thank J.-J. Yang for his contribution at the earlier stage of this work. 
We are grateful to W. Vogelsang and M. Stratmann for an helpful correspondance
and useful comments.
\end{acknowledgments}

\newpage
\end{document}